\documentclass[10pt]{article}
\usepackage{amsmath, amsthm, amssymb, bm}
\usepackage{amsfonts}
\usepackage{amscd}
\usepackage{delarray}
\usepackage{graphicx}
\newcommand{\bra}[1]{\langle #1|}
\newcommand{\ket}[1]{|#1\rangle}
\newcommand{\braket}[2]{\langle #1|#2\rangle}
\newcommand{\Tr}{\mathrm{Tr}}

\begin{document}

\title{The preparation time in a scattering experiment}
\author{P Bryant\footnote{CCQS, Department of Physics, University of Texas at Austin, Austin, TX 78712}}

\maketitle
\begin{abstract}
A quantum mechanical theory with time asymmetry intrinsic to states (or observables)
features the concept of an initial time of the state and thus a preparation
time of the physical system represented by the state.
This special time is investigated in the context of scattering theory, where, 
in standard quantum mechanics, the physical meaning of a preparation time has remained obscure.
In an experiment, the preparation time corresponds to an ensemble of times of scattering
marking the times in the laboratory when one scattering projectile interacts with one target quantum.
\end{abstract}

\section{Introduction}
Standard quantum mechanics has time evolution for states (in the Schr\"odinger
picture) or for observables (in the Heisenberg picture) that extends from $-\infty<t<\infty$.
It therefore makes predictions for experimental (Born) probabilities also for
$-\infty<t<\infty$.
Such predictions are not intuitive because, in an experiment, an observable can be measured
in a system represented by a state only after the system has
been prepared at some finite time.
There exists, however, a causal quantum theory,
which makes predictions of Born probabilities only for times
$t\ge0$, where $t=0$ corresponds to the time at which the 
quantum system has been prepared \cite{fortschritte_der_physik_1}.

To determine if the concept of a preparation time incorporates naturally
into quantum mechanical scattering theory,
one must understand what this special time represents phenomenologically.
Section \ref{arrow_of_time} is a brief motivation for a theory that addresses the preparation
time of physical systems.
Section \ref{experiments} is a discussion of the non-relativistic scattering cross section and
its relation to the preparation time.
In Section \ref{ensemble},
the preparation time of systems represented by scattered states is identified for a scattering experiment.

\section{Conjecturing an Arrow of Time}
\label{arrow_of_time}
For simplicity, let us discuss systems represented by pure states described by a state vector $\phi$
(or density operator $\rho=\ket{\phi}\bra{\phi}$.)
Let the observable be represented by a vector $\psi$
(or the corresponding observable $\Lambda=\ket{\psi}\bra{\psi}$.)
In the theory, the time evolution of these quantum mechanical entities is given for states
by the Schr\"odinger equation
\begin{equation}
\label{schrod_eqn}
i\hbar\frac{\partial}{\partial t}\phi=H\phi
\end{equation}
and for observables by the Heisenberg equation
\begin{equation}
\label{heis_eqn}
i\hbar\frac{\partial}{\partial t}\psi=-H\psi,\quad\textrm{or}\quad
i\hbar\frac{\partial}{\partial t}\Lambda(t)=[\Lambda(t),H].
\end{equation}
In standard quantum mechanics,
one uses as a boundary condition for these dynamical equations the Hilbert space
axiom:
\begin{equation}
\label{hilbert_space_bc}
\{\phi\}=\{\psi\}=\textrm{Hilbert space}=\mathcal{H}.
\end{equation}
Then, by the Stone-von Neumann theorem \cite{stone_1932,von_neumann_1932}, the solutions of the dynamical equations
(\ref{schrod_eqn}) or (\ref{heis_eqn}) are
\begin{equation}
\label{state_evolution}
\phi(t)=e^{-\frac{i H t}{\hbar}}\phi_0\,,\qquad-\infty<t<\infty,
\end{equation}
for states, or
\begin{equation}
\label{obs_evolution}
\psi(t)=e^{\frac{i H t}{\hbar}}\psi_0\,,\qquad-\infty<t<\infty,
\end{equation}
for observables.
This is time evolution given for states by the unitary group
\begin{equation}
\label{state_group}
U^\dagger(t)=e^{-\frac{i H t}{\hbar}}\,,\qquad -\infty<t<\infty,
\end{equation}
and for observables by the unitary group
\begin{equation}
U(t)=e^{\frac{i H t}{\hbar}}\,,\qquad -\infty<t<\infty.
\end{equation}
The group product is
\begin{equation}
U^\dagger(t_1)U^\dagger(t_2)=U^\dagger(t_1+t_2).
\end{equation}
For every evolution $U(t)$ there exists the inverse
\begin{equation}
U^{\dagger}(t)^{-1}=U^\dagger(-t).
\end{equation}
The unitary group evolution of states (or 
of observables) is symmetric in time; one can use any value of the time parameter,
$-\infty<t<\infty$, to drive
the evolution of a state (or, in the Heisenberg picture, of an observable) forward or backward.

A reversible time evolution for the physical systems \textit{represented} by
states is not intuitive or observed, but the reversible time evolution
\eqref{state_evolution} or \eqref{obs_evolution} is generally accepted.
In a quantum mechanical experiment, the calculated, measurable quantity is
(in the Schr\"odinger picture)
the probability to
find the observable $\Lambda=\ket{\psi}\bra{\psi}$ in the time-varying state $\phi(t)$.
This quantity is the Born probability:
\begin{equation}
\label{born_probability}
\mathcal{P}_{\phi(t)}(\Lambda)=\Tr\big(\Lambda\ket{\phi(t)}\bra{\phi(t)}\big)=\Tr\big(\ket{\psi}\braket{\psi}{\phi(t)}\bra{\phi(t)}\big)=|\braket{\psi}{\phi(t)}|^2.
\end{equation}
Note that the Born probability to find the observable $\Lambda$ in the state $\phi(t)$
has time evolution given by the state $\phi(t)$.
Comparing (\ref{born_probability}) with (\ref{state_evolution}) and (\ref{state_group}), one finds
\begin{equation}
\label{reversible_born}
\mathcal{P}_{\phi(t)}(\Lambda)=|\braket{\psi}{\phi(t)}|^2=|\braket{\psi}{U^\dagger(t)\phi_0}|^2,\qquad -\infty<t<\infty.
\end{equation}
Again, one can use any value between $-\infty$ and $\infty$ for the time parameter.
One has not only reversible time evolution for states, but also reversible time evolution
for the predictions of the experimentally measured quantities.

Phenomenologically, however, one knows that a measurement cannot be made before
an experimental system has been prepared.
In terms of the theoretical objects,
an observable $\Lambda$ cannot be measured
in the state $\phi(t)$ before the system represented by the state
$\phi(t)$ has been prepared at some finite time $t_{prep}>-\infty$.
This is a statement of the preparation-registration arrow of time,
and it emphasizes the notion of a finite preparation time.
In time symmetric quantum mechanics,
the theoretical predictions of experimental quantities \eqref{reversible_born}
also hold for times $-\infty<t<t_{prep}$.

A quantum theory providing asymmetric time evolution for states, which matches
the phenomenologically observed, asymmetric time evolution of physical systems,
and having as a feature
the preparation time, has been obtained \cite{fortschritte_der_physik_1,bohm_antoniou_kiel_arrow_of_time}.
By changing the boundary conditions \eqref{hilbert_space_bc} for the dynamical equations
of quantum mechanics,\footnote{
Specifically, one differentiates mathematically between in-states, $\phi^+$,
defined by a preparation apparatus (accelerator),
and out-observables, $\psi^-$, defined by a registration apparatus (detector) \cite{bohm_early_theory_1981,bohm_early_theory_1978}.
The sets of energy wave functions for in-states and out-observables are chosen to be
$\{\braket{^+E}{\phi^+}\}=(H^2_-\cap\mathcal{S})\vert_{\mathbb{R}_+}$ and
$\{\braket{^-E}{\psi^-}\}=(H^2_+\cap\mathcal{S})\vert_{\mathbb{R}_+}$, respectively, 
where $H^2_\mp$ are the Hardy function spaces, and $\mathcal{S}$ is the Schwartz space \cite{gadella_hardy_rhs}.}
thus modifying
the axiom \eqref{hilbert_space_bc} of standard quantum theory,
one finds the time evolution to be \cite{fortschritte_der_physik_1,bohm_annals_of_physics_small_preprint}
\begin{equation}
\label{asym_states}
\phi(t)=e^{-\frac{i H^\times t}{\hbar}}\phi_0\,,\qquad 0\leq t<\infty,
\end{equation}
for states or
\begin{equation}
\label{asym_observables}
\psi(t)=e^{\frac{i H t}{\hbar}}\psi_0\,,\qquad 0\leq t<\infty,
\end{equation}
for observables.\footnote{
Technically, the time in (\ref{asym_states}) or (\ref{asym_observables}) is limited
from below by some finite time $\tilde t$:
$t\geq \tilde t > -\infty$.  Let us choose $\tilde t=0$.
}
This is no longer time evolution given by the unitary group $U^\dagger(t)$,
but is rather given by a \textit{semigroup},
\begin{equation}
\label{semigroup_states}
U^\times(t)=e^{-\frac{i H^\times t}{\hbar}}\,,\qquad 0\leq t<\infty,
\end{equation}
which is the time evolution operator for states, and by another \textit{semigroup},
\begin{equation}
\label{semigroup_observables}
U(t)=e^{\frac{i H t}{\hbar}}\,,\qquad 0\leq t<\infty,
\end{equation}
which is the time evolution operator for observables.\footnote{
The operator notation $A^\times$ signifies that the operator is an extension of a Hilbert space
selfadjoint operator $A=A^\dagger$ onto the space dual to the space of vectors describing states
and observables.}
While the group product is still defined,
\begin{equation}
U^\times(t_1)U^\times(t_2)=U^\times(t_1+t_2),
\end{equation}
being a semigroup means that the inverse, $U^{\times}(t)^{-1}$, of an element,
$U^\times(t)$ with $t>0$, does not exist.
Simply stated, one cannot evolve the state vector representing a physical system
backward in time.
The Born probability is calculated for $0\leq t<\infty$ only:
\begin{equation}
\label{asymmetric_born}
\mathcal{P}_{\phi(t)}(\Lambda)=|\braket{\psi}{\phi(t)}|^2=|\braket{\psi}{U^\times(t)\phi_0}|^2,\qquad 0\leq t<\infty.
\end{equation}
As will be discussed in the next section, one identifies the time $t=0$ in (\ref{asymmetric_born})
as the preparation time of the physical system represented by the state vector $\phi(t)$.
Therefore, one no longer has a Born probability for finding an observable in a state
before the system represented by that state is prepared.
With the time asymmetric quantum theory,
calculations of experimental quantities naturally incorporate the preparation time.

\section{Scattering Experiments}
\label{experiments}
The asymmetric time evolution provided by the semigroup operators (\ref{semigroup_states})
and (\ref{semigroup_observables}) has been
called \textit{intrinsic} time asymmetry \cite{physica_a_1997,bohm_antoniou_kiel_arrow_of_time}, to distinguish it from the extrinsic time asymmetry
investigated for open quantum systems \cite{davis_open_systems,petruccione_open_quantum_systems}.
Because the Born probability (\ref{asymmetric_born}) exists only for time $t\geq 0$,
the time $t=0$ is the time when an observable is just ready to be detected.
Because a system described by the state $\phi(t)$ must be prepared
before an observable can be detected in it,
one identifies the time $t=0$ as the preparation time of that microphysical system described
by the state $\phi(t)$.
At its moment of preparation, the system is represented by the state vector at $t=0$: $\phi(t=0)$.

For a scattering experiment, one has a detector,
represented by the observable $\Lambda=\ket{\psi^-}\bra{\psi^-}$,
which is built to detect that observable of a scattered, microphysical system represented
by a scattered state called $\phi^+(t)$.
The experimentally observable quantity is the differential cross section, which is calculated as
\begin{equation}
\label{cross_section}
d\sigma(\vartheta,\varphi)=
\frac{\left(\begin{array}{l}\textrm{transition probability per unit time} \\ \textrm{for scattering quanta into }\Delta\Omega\end{array}\right)\Big|_{t=0}}
{\left(\begin{array}{l}\textrm{incident probability per} \\ \textrm{unit time and unit area} \end{array}\right)}.
\end{equation}
Here, $\Delta\Omega$ is the solid angle subtended by the detector.

One calculates the transition probability rate in \eqref{cross_section} at the time
$t=0$ \cite{gell-mann_goldberger_formal_theory_scattering_1953}.
The transition probability is given, according to \eqref{asymmetric_born}, by 
\begin{equation}
\label{scattering_born}
\mathcal{P}_{\phi^+(t)}\big(\ket{\psi^-}\bra{\psi^-}\big)=|\braket{\psi^-}{\phi^+(t)}|^2,\qquad 0\leq t<\infty.
\end{equation}
If one is to identify the time $t=0$ in \eqref{cross_section}
as the semigroup time $t=0$ in \eqref{scattering_born},
as seems natural,
then the transition probability rate in \eqref{cross_section} is calculated
precisely at the \textit{initial} time of a scattered state, which
coincides with the preparation time of a scattered, microphysical system.
The initial time of a scattered state is therefore the time of scattering of the
system it represents.

A consequence of the existence of a preparation time
is the necessary, theoretical distinction between the time associated with
a state $\phi^+(t)$ and the time associated with any external state or reservoir
with which $\phi^+(t)$ might interact.
The time $t=0$ is the semigroup time of \eqref{semigroup_states},
which is associated with a particular state,
and it is independent of the time parameterizing the evolution of any external state.
For the purpose of discussion, take the external state for a scattering experiment
to be the laboratory.
Even in the non-relativistic case, one distinguishes between
the microscopic time $t$ of $\phi^+(t)$ and the macroscopic time marked by the clocks on the wall of the laboratory, $t_{lab}$.
It is only the microscopic time, belonging to the state $\phi^+(t)$
(or, in the Heisenberg picture, to the observable $\psi^-(t)$) that is bounded by $t\geq 0$.

\section{Ensemble of Times}
\label{ensemble}
Consider the theoretical description of a scattering experiment \cite{taylor_scattering,newton_scattering_theory_book}.
The controlled in-state vector, called $\phi^{in}(t)$,
represents the incoming, projectile system prepared, up to phase,
by an accelerator.
The uncontrolled out-state vector, $\phi^+({t})$, represents a
microphysical, scattered system, and it is 
defined by the controlled in-state 
as well as by the scattering interaction.
This microphysical, scattered system
is prepared at the time of scattering, which is the time of beam crossing.\footnote{
``time of beam crossing'' here refers to the moment a single bunch of projectile particles arrives at the
target (fixed-target experiment) or to the moment when a bunch of projectiles moving one direction
crosses a bunch moving in the opposite direction (collider experiment).}
Typical scattering experiments consist of multiple beam crossings occurring over
a span of days or years, as counted by the macroscopic time of the lab.

On the microphysical level, the uncontrolled out-state represents a scattered system with
characteristics specified by the vector $\phi^+({t})$.
One does not distinguish between two instances of that state (two different microphysical systems),
prepared in an experiment, if the only difference between them is the macroscopic
laboratory time, $t_{lab}$, of scattering.
In other words, if a scattered, microphysical system is prepared during a beam crossing today,
it is considered phenomenologically identical to one prepared during an identical beam crossing yesterday.
These separate yet phenomenologically identical instances of the state (microphysical systems) are
all represented by $\phi^+(t)$.

Furthermore, every state $\ket{\phi^+(t_1)}$ evolves over time into a unique state
$\ket{\phi^+(t_2)}$:
\begin{equation}
\ket{\phi^+(t_2)}=U^\times(t_2-t_1)\ket{\phi^+(t_1)},\quad t_2>t_1.
\end{equation}
One can choose to describe two different instances of the state $\phi^+(t)$
by two different vectors $\phi_a^+$ and $\phi_b^+$.
However, if one wishes the two microphysical systems, and thus the two vectors,
to be equivalent at some later time $t_{meas}$
when a measurement is to be made,
then by the linearity of the unitary time operators \eqref{semigroup_states}
\begin{eqnarray}
\ket{\phi_a^+(t_{meas})}-\ket{\phi_b^+(t_{meas})}=&0&=U^\times(t_{meas})\ket{\phi_a^+}-U^\times(t_{meas})\ket{\phi_b^+} \nonumber \\
 & & =  U^\times(t_{meas})\big(\ket{\phi_a^+} -\ket{\phi_b^+} \big).
\end{eqnarray}
Therefore, $\ket{\phi_a^+}=\ket{\phi_b^+}$, and the two
different scattering systems are in fact represented by the same state.\footnote{
If the microphysical system represented by $\phi_a^+$ is
prepared, say, at a \textit{microphysical} time $\Delta t$ later than
the system represented by
$\phi_b^+$ is prepared, the same argument gives
$\ket{\phi_a^+}=U^\times(\Delta t)\ket{\phi_b^+}$, and the same conclusion follows.
Note that this microphysical (though finite) time $\Delta t$ is independent of the
macroscopic times in the lab at which the systems represented by $\phi_a^+$ and
$\phi_b^+$ were prepared.
}
It follows that the state $\phi^+(t)$ represents
an ensemble of microphysical, scattered systems.

The ensemble of systems contains members prepared during 
various beam crossings at various macroscopic times in the lab.
One is free to consider the macroscopic time in the lab at which a given member of the ensemble was prepared, or scattered.
This macroscopic time would correspond to a specific time of beam crossing, and to the ensemble
of \textit{microphysical} systems would belong an ensemble of \textit{macrophysical} times
of scattering.
On the microphysical level, however,
every member of the ensemble represented by $\phi^+(t)$ is prepared at
the microphysical time $t=0$.

In the non-relativistic case, one can relate the microscopic and macroscopic times quite easily.
Let $t^i$ denote the macroscopic time corresponding to the $i$-th member of the ensemble.
Then $t_0^i$ would be the macroscopic time of the preparation of the $i$-th member of the ensemble.
One can write
\begin{equation}
t=t^i-t_0^i.
\end{equation}
The microphysical time is $t$.
The ensemble of macroscopic preparation times is  $\{t_0\}$.

Physicists preparing data from a scattering experiment are aware of the ensemble,
which is an intuitive notion.
Experimental results are gathered from an ensemble of events occurring over a period of days or years.
Of course, quantum mechanical calculations result in probabilities, and they do not
address individual microphysical systems, which are particular instances of a state.
The individual times within the ensemble $\{t_0\}$ are not reproducible.

\section{Conclusion}
The time asymmetric theory of quantum mechanics includes the
theoretical notion of a special time, $t=0$, corresponding to the
preparation time of a system represented by a state.
This time incorporates naturally into the description of scattering experiments.
Its existence emphasizes the theoretical
difference between time belonging to a scattered state, $\phi^+(t)$, and the time belonging
to the state of an external system such as the laboratory.
Phenomenologically, the preparation time corresponds to an ensemble of macroscopic \textit{times of scattering}
marking the time in the laboratory when one interaction event
between one projectile and one target quantum occurs.

\section*{Acknowledgments}
The author would like to thank the organizers of the Fifth International Symposium on
Quantum Theory and Symmetries, and in particular Manuel Gadella and
Mariano del Olmo, for their kind hospitality.
The author is also grateful for the advice and support from Arno Bohm and Yoshihiro Sato.

\bibliographystyle{plain}
\bibliography{/home/pbryant/Documents/SMWork/QM/qm_bibliography}

\end{document}